\documentclass{ws-procs9x6}

\usepackage{amsmath,amssymb,graphics,graphicx,epsfig}

\begin{document}

\title{Phenomenology from a $U(1)$ guaged hidden sector\footnote{
\uppercase{R}esearch supported in part by the \uppercase{N}ational 
\uppercase{S}cience and \uppercase{E}ngineering \uppercase{C}ouncil of 
\uppercase{C}anada, and by \uppercase{T}aiwan \uppercase{NSC} grant 
95-2112-M-007-032.}}

\author{W. F. Chang}

\address{Department of Physics, National Tsing-Hua University, \\
Hsinchu 300, Taiwan}

\author{J. N. Ng and J. M. S. Wu\footnote{\uppercase{S}peaker at the 
\uppercase{L}ake \uppercase{L}ouise \uppercase{W}inter \uppercase{I}nstitute 
2007, \uppercase{F}eb. 19-24, \uppercase{A}lberta, \uppercase{C}anada.}}

\address{Theory group, TRIUMF, \\ 
4004 Wesbrook Mall, \\
Vancouver, BC, V6T 2A3, Canada}  

\maketitle

\abstracts{
We consider the phenomenological consequences of a hidden Higgs sector 
extending the Standard Model (SM), in which the matter content are uncharged 
under the SM gauge groups. We consider a simple case where the hidden sector 
is gauged under a $U(1)$ with one Higgs singlet. The only couplings between SM
and the hidden sector are through mixings between the neutral gauge bosons of 
the two respective sectors, and between the Higgs bosons. We find signals 
testable at the LHC that can reveal the existence and shed light on the nature
of such a hidden sector.}

\section{The shadow $U(1)_s$ model}
It has been recently pointed out that hidden sectors which commonly extend the
Standard Model (SM), need not be associated with a very high energy scale, and
renormalizable interations with the SM fields through mixing are possible which
provide portals to new physics accessible at the Large Hadron Collider 
(LHC)~\cite{PW06}.

We consider here a simple case where the hidden sector contains a single 
complex scalar $\phi_s$ gauged under the hidden sector gauge group, which we 
take to be a single ``shadow'' $U(1)_s$. The complete Lagrangian of our model 
takes the form~\cite{CNW06}
\begin{equation}\label{Eq:Lagrangian}
\mathcal{L} = \mathcal{L}_{SM}
-\frac{1}{4}X^{\mu\nu}X_{\mu\nu}-\frac{\epsilon}{2}B^{\mu\nu}X_{\mu\nu}
+\left|\left(\partial_{\mu}-\frac{1}{2}g_s X_{\mu}\right)\phi_s\right|^2
-V_0(\Phi,\phi_s) \,,
\end{equation}
where $B^{\mu\nu}$ and $X^{\mu\nu}$ are the field strength tensors of the SM
$U(1)_Y$ and $U(1)_s$ respectively, $\Phi$ is the SM Higgs field, and $g_s$ is
the gauge coupling constant of the $U(1)_s$. The tree level scalar potential 
is given by
\begin{equation}\label{Eq:V0}
V_0(\Phi,\phi_s) =
\mu^2\Phi^\dag\Phi + \lambda(\Phi^\dag\Phi)^2 + 
\mu_s^2\phi_s^*\phi_s + \lambda_s(\phi_s^*\phi_s)^2 +
2\kappa\left(\Phi^\dag\Phi\right)\left(\phi_s^*\phi_s\right) \,.
\end{equation}
The hidden sector couples to the SM only through the two mixing terms, the 
kinetic mixing between the two $U(1)$'s parameterized by $\epsilon$, and the 
mixing between the scalar fields controlled by $\kappa$. 

The spontaneous symmetry breaking (SSB) of the symmetry
$SU(2)_L \times U(1)_Y \times U(1)_s$ down to $U(1)_{EM}$ is triggered once 
the scalars acquire nonzero VEVs:
\begin{equation}
\langle\Phi\rangle = \frac{1}{\sqrt{2}}
\begin{pmatrix}
0 \\
v
\end{pmatrix} \,, \quad
\langle\phi_s\rangle = \frac{v_s}{ \sqrt{2}} \,.
\end{equation}

\section{Mixing in the gauge sector}
Because of the kinetic mixing term, a $GL(2)$ transformation is needed to 
recast the Lagragian in Eq.~\eqref{Eq:Lagrangian} to canonical form, which
mixes the gauge fields of the $U(1)_Y$ and $U(1)_s$: 
\begin{equation}
\begin{pmatrix}
X \\
B
\end{pmatrix} 
=
\begin{pmatrix}
 c_\epsilon & 0 \\
-s_\epsilon & 1
\end{pmatrix} 
\begin{pmatrix}
X' \\
B'
\end{pmatrix} \,, \quad 
s_\epsilon = \frac{\epsilon}{\sqrt{1-\epsilon^2}} \,,\quad
c_\epsilon = \sqrt{1-s_\epsilon^2} \,.
\end{equation}
A further mass mixing happens after SSB between the SM $Z$ and the extra 
``shadow'' $Z_s$ bosons, with the mixing angle given by
\begin{equation}
\tan(2\eta) = \frac{2s_W s_\epsilon}{c_W^2(M_3/M_W)^2+s_W^2 s_\epsilon^2-1} 
\,,\quad M_3 = \frac{g_s v_s}{2} \,,
\end{equation}
where $s_W$ denotes the weak-mixing angle $\sin\theta_W$, 
$c_W = \sqrt{1-s_W^2}$, and $M_W = g_W v/2$ is the $W$ mass with $g_W = e/s_W$.

\begin{figure}[htbp]
\centering
\includegraphics[width=2.8in]{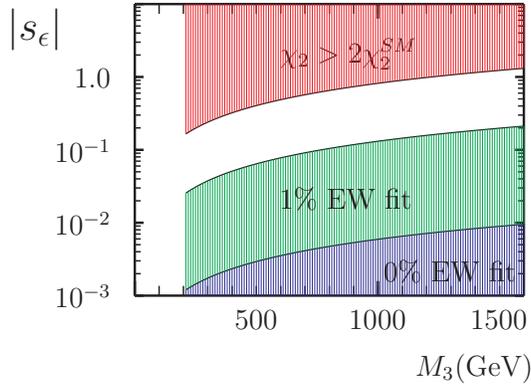}
\caption{\label{fig:GFit_BM3_fig}
The bound on $s_\epsilon$ and $M_3$ from EWPTs. The upper band is region 
excluded by too large a deviation from the SM, $\chi_2 > 2 \chi_2^{SM}$. The 
middle one is the allowed region where $(\triangle\chi_2/\chi_2^{SM})<0.01$. 
The lower band is the region where the global fit gives comparable results to 
the SM.}
\end{figure}

These mixings modify couplings of $Z$ and introduce new ones to $Z_s$ which
directly affect electroweak precision tests (EWPTs) that stringently constrain
any model with extra $Z$ bosons, which in turn constrain the kinetic mixing 
parameter $\epsilon$. The results of a systematic study of all the currently 
available EWPT observables are summarized by Fig.~\ref{fig:GFit_BM3_fig}.
Here, $\chi_2(s_\epsilon,\,M_3)$ measures the deviation between the model and 
the experiments, and $\triangle\chi_2\equiv\chi_2-\chi_2^{SM}$ with 
$\chi_2^{SM}\equiv\chi_2(0,\,M_3) = \chi_2(s_\epsilon,\,\infty)$.~\footnote{
See~\cite{CNW06} for more details.}

As seen from Fig.~\ref{fig:GFit_BM3_fig}, $\epsilon$ need not be vanishingly
small as is usually assumed; it can be of order $10^{-3} \sim 10^{-2}$, in 
agreement with the general expectation from string theory~\cite{string}.

\section{$Z_s$ signal at the LHC}
The phenomenology of the $Z_s$ is expected to be very different from scenarios
where the extra $Z$ couples directly to the SM, such as in the familiar 
$SO(10)$ or $E_6$ based grand unified theories (GUTs) models. One immediate
example is the narrowness of the $Z_s$ width. In the large $Z_s$ mass limit,
say $M_{Z_s} > 1$~TeV,
\begin{equation}
\Gamma_{Z_s} \simeq 2.37\frac{g_2^2 M_{Z_s}s_\epsilon^2}{24\pi c_W^2}
= 0.1742\left(\frac{M_{Z_s}}{1\,\mathrm{TeV}}\right)
\left(\frac{s_\epsilon^2}{0.01}\right)\mathrm{GeV} \,.
\end{equation}

\begin{figure}[htbp]
\centering
\includegraphics[width=3in]{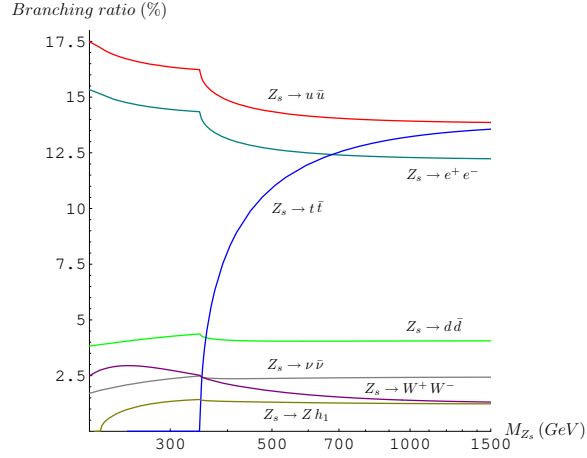}
\caption{\label{fig:ZpBr_fig}
Branching ratio for the $Z_s$ decays as functions of $M_{Z_s}$. The mass of the
Higgs is taken to be $M_{h_1}=120$~GeV. The parameter $s_\epsilon$ is set to
be $10^{-3}$.}
\end{figure}

Another distinguishing feature is the $Z_s$ branching ratios, as shown in
Fig.~\ref{fig:ZpBr_fig}.
The $Z_s$ decays preferentially into $u$-type quarks and charge leptons,
which is very different from the SM $Z$ decay. Also for a sufficiently heavy 
$Z_s$, the branching ratio into charge leptons and $t$ quarks is relatively 
large and almost equal. This can be used to distinguish between different extra
$Z$ models and may also be used as a diagonstic tool at the LHC.

\section{A classically conformal Higgs sector}
Motivated by the idea that very light hidden sector scalars may be candidates
for dark matter~\cite{BF04}, we consider here a special case where our model
is classically conformal (setting $\mu = \mu_s = 0$ in Eq.~\ref{Eq:V0}). The 
SSB is induced radiatively via Coleman-Weinberg (CW) mechanism~\cite{CW73}, 
which naturally generates a small mass scale without further assumption or 
fine tuning.

Applying the perturbative multiscalar effective potential analysis of Gildener
and S. Weinberg~\cite{GW76}, two physical scalar states arise. One is a heavy
SM-like Higgs boson, $H_2$, and the other a light ``shadow'' Higgs, $H_2$, 
whose mass arise entirely from radiative corrections and is given
 by~\cite{CNW07}
\begin{equation}\label{Eq:mssqse0}
M_{H_1}^2 = 
\frac{3v_r^2}{64\pi^2(1 + r)}\left[\frac{3g_W^4}{2} + g_Y^2 g_W^2 +
\frac{g_Y^4}{2} + \frac{8M_{Z_s}^4}{v_r^4}\right] + 
\frac{M_{H_2}^4-12M_t^4}{8\pi^2 v_r^2(1 + r)} \,,
\end{equation}
where $r\equiv\sqrt{\lambda/\lambda_s} = 4M_{Z_s}^2/(v_r^2 g_s^2)$, and
$v_r \equiv v/\sqrt{1+r} = 4M_W^2/g_W^2$ is fixed by the physical $W$ mass.

\section{Search for the light shadow Higgs at the LHC}
The Yukawa couplings of the shadow Higgs to the SM fields is simply that of
the SM Higgs scaled by a factor of $1/\sqrt{1+r^2}$. Applying the bounds from 
the LEP direct Higgs search to the shadow Higgs case, which is most stringent
at $M_{H_1} \simeq 20$~GeV~\cite{LEP2}, we have 
$\xi^2\equiv(g_{HZZ}/g_{HZZ}^{SM})^2 = 1/(1+r) \lesssim 2 \times 10^{-2}$ 
implying that $r \gtrsim 49$. From the expression of $r$, this bound can be 
easily satisfied for appropriate choices of $M_{Z_s}$ and $g_s$, and a light 
shadow Higgs is not ruled out.

Since the shadow Higgs couples like the SM Higgs, one way to search for it at 
the LHC, is to studying the $t \rightarrow H_1 b\,W^+$ decay just like for the
SM Higgs. Suppose $M_{H_1} = 30$~GeV, taking the top-Higgs Yukawa coupling to 
be $y_t \sim 1$, the decay width is 
\begin{equation}
\Gamma(t \rightarrow H_1 b\,W^+) \sim \frac{2 \times 10^{-3}}{1+r} \,
\mathrm{GeV} \,.
\end{equation}
This is to be compared with that in the SM, 
$\Gamma_t^{SM} = 1.37$~GeV~\cite{JK89}. With $r \gtrsim 49$, a search for the 
shadow Higgs in the $t \rightarrow H_1 b\,W^+$ decay is likely to require the 
LHC to operate at high luminosity for extended periods of time.

\section{Summary}
Renormalizable mixing between the hidden and the SM sectors are portals through
which new physics can be discovered using the LHC. One distinct signature of a 
hidden $U(1)$ sector is the existence of an extra $Z$ with a very narrow width.
To distinguish it from that of the other extra $Z$ models, precise measurement
of its branching ratios is needed, although the International Linear Collider
would provide a much cleaner environment for doing so than the LHC.

In the special case where our model is classical conformal, a light shadow 
Higgs can be generated from the SSB of the scale-invariance through CW 
mechanism. It is viable under the current direct search limit, and can be 
searched for at the LHC in the $t \rightarrow H_1 b\,W^+$ decay. However, to 
achieve the required detection sensitivity, high luminosity runs would likely 
be needed.

\end{document}